\documentstyle[epsf,preprint,aps]{revtex}
\draft
\preprint{SNUTP 00-017}
\begin{document}
\title{\Large\bf 
The radion contribution to the weak mixing angle}
\author{Jihn E. Kim \footnote{jekim@phyp.snu.ac.kr}, 
Bumseok Kyae \footnote{kyae@fire.snu.ac.kr}, and
Jong Dae Park \footnote{jdpark@phya.snu.ac.kr}} 
\address{ Department of Physics and Center for Theoretical
Physics, Seoul National University,
Seoul 151-742, Korea}
\maketitle

\begin{abstract} 
In the Randall-Sundrum (RS) compactification,
the gap between two branes is stabilized by
the vacuum expectation value of a scalar field called
{\it radion}, $\phi$. This radion behaves like
a weak interaction singlet scalar field, coupling
to matter through the trace of the energy momentum
tensor. We find that it can induce a sizable
correction to the weak mixing angle if the 
vacuum expectation value and the mass of the radion
is around TeV. We also comment on the contribution to
the $K_L-K_S$ mass difference and to $(g-2)_\mu$. 
\end{abstract}

\newpage

The recent try of the brane physics on an $S^1/Z_2$ orbifold model 
with non-factorizable geometry of space-time 
\cite{rs} has attracted a great deal of attention.    
It is due to a possibility of generating a large hierarchy
of mass scales between two branes, 
Brane 1 (B1) with a positive cosmological
constant(or brane tension) $\Lambda_1\equiv 6k_1M^3$ and Brane 2 (B2) 
with a negative cosmological constant $\Lambda_2\equiv 6k_2M^3$.
The bulk between these branes is required to carry a negative 
bulk cosmological constant $\Lambda_b\equiv -6k^2M^3$. 
B1 is interpreted as the hidden brane with a
fundamental mass scale and B2 is identified with the visible brane.
In this setting the metric at B2 has an exponential warp factor which could be
used to understand the huge gap between the Planck and eletroweak scales.
Although this model introduces cosmological constants $k$ in the bulk and
$k_1$ and $k_2$ on the branes, it still describes a static universe because of
the fine-tuning between the bulk and brane cosmological constants 
$k=k_1=-k_2$, which are the consistency conditions in the model.  

Hence, if the fine-tuning is not
exact, the solution has the time dependence and the universe
expands exponentially~\cite{nihei} but its form is not suitable for 
the standard Big Bang universe after the inflation. To circumvent this 
cosmological problem, some approximation schemes 
regarding the brane matter in the static limit has been taken into
account and(or) some conditions (such as the positive
brane tension) for the brane and bulk cosmological constants are 
required~\cite{cline}. With the addition of the Gauss-Bonnet
interaction, one can have a finite region of parameter
space allowing a positive brane tension at B2~\cite{kkl}.
In Ref.~\cite{cline}, B1 was considered as the visible brane  
just to circumvent the cosmological difficulty, which contradicts
the original motivation for the gauge hierarchy solution~\cite{rs}. 
Since this problem can be resolved with 
the Gauss-Bonnet interaction \cite{kkl}, bulk matter effects and 
extra dimension stabilization process \cite{rs3}, etc.,
we consider B2 as the visible brane in this paper.

The simplest example is in the five dimensional world with
the fifth dimension denoted as $y$.
Upon compactification, the $(yy)$-component of the metric
tensor (a moduli) behaves like a scalar field 
which can be light ($\sim$ electroweak scale). This
moduli is often called the {\it radion} since its
vacuum expectation value determines the distance scale
between B1 and B2. Since the radion can be light,
it may have detectable signatures at the present
and future accelerator experiments if the vacuum
expectation value of the radion is as small
as the electroweak scale \cite{radion}. In this scenario, we
study the radion contribution to the W and Z masses or
to the correction to the weak mixing angle $\sin^2\theta_w$.
In addition, we comment briefly on the radion contribution to
the other parameters of the electroweak physics, but they
do not give very strong constraints.
 
The fields fluctuating near the RS background are given as 
\begin{equation}
\label{eq:fluc}
ds^2 = e^{- 2 kb(x)|y|} g_{\mu\nu}(x) dx^\mu dx^\nu - b^2(x)dy^2,
\end{equation}
where $g_{\mu\nu}$ is the four-dimensional graviton and $T(x)$ is the modulus 
field.  The $S^1/Z_2$ symmetry forbids the gravi-photons $g_{\mu 5}$. 
After the Kaluza-Klein (KK) reduction for the massless modes, 
the 5-dimensional Hilbert-Einstein action gives \cite{gwnew}  
\begin{equation}
S= \frac{M^3}{2k} \int d^4x \sqrt{-g} \left(1-e^{-2 kb(x)}\right)R 
+ \frac{3M^3}{k}\int d^4 x \sqrt{-g}
\partial_\mu \left(e^{-kb(x)}\right) \partial^\mu \left(e^{-kb(x)}\right)~,
\end{equation}
where $R$ is the 4-dimensional Ricci scalar.
Let us take field definition as follows,  
\begin{equation}\label{4dscalar}
\phi(x)=\langle\phi\rangle+\varphi(x)
\equiv t~e^{-kb(x)}~~ {\rm with} ~~t=\sqrt{6M^3/k}~, 
\end{equation}
where $\langle\phi\rangle$ is a vacuum expectation value (VEV) of $\phi(x)$ 
and $\varphi(x)$ is its fluctuation near the VEV.  
Note that the $t$ has mass dimension of the Planck scale.   
Then, we arrive at
\begin{equation}
\label{action}
S= \frac{2 M^3}{k} \int d^4 x \sqrt{-g}
\left(1-(\langle\phi\rangle+\varphi)^2/t^2\right) R 
+\frac{1}{2}\int d^4 x\sqrt{-g}\partial_\mu \varphi\partial^\mu\varphi~,
\end{equation}
from which we see the modulus field $b(x)$ can be interpreted as 
a 4-dimensional scalar field.  
We will call the scalar $\varphi$ ``radion" \cite{radion}.  
The radion is basically massless because it is associated with the metric 
component. But it could get mass and vacuum expectation value 
after KK reduction. For example, 
the Goldberger-Wise mechanism \cite{gw} or a gaugino condensation  
in supergravity models would render these to the radion. Therefore,
we will assume its mass and vacuum expectation value below.

To obtain radion's phenomenological constraints, one must
derive its couplings to gauge bosons and to fermions.
The kinetic energy terms of the vector boson and fermion fields, 
which live on the visible ($y=1/2$) brane, read 
\begin{equation}
{\cal L} = e\bigg[
-\frac{1}{4}g^{MO}g^{NP}F_{MN}F_{OP}
+\frac{i}{2}\left(\overline{\psi}\gamma^{\mu}e^{M}_{\mu}\nabla_M\psi
-e^{M}_{\mu}(\nabla_M\overline{\psi})\gamma^{\mu}\psi\right)\bigg]~,
\end{equation} 
where $\nabla$ denotes the covariant derivative on a curved manifold, 
and $e^{M}_{\mu}$ and $e$ are the inverse and determinant of the vierbein,  
respectively.  
Here $M$, $N$ indicate generally Einstein indices of the curved space 
and $\mu$ is the Lorentz index of the local reference frame, but in our case 
the distiction between them is meaningless because our space is conformally 
flat.  
Under the transformation 
\begin{eqnarray}\label{vierbeintrf}
e^{\mu}_{M}&&\longrightarrow e^{-kb}e^{\mu}_{M}\approx e^{-kb}\delta^{\mu}_{M},
~~~~~{\rm or}\nonumber \\
g_{MN}&&\longrightarrow 
e^{-2kb(x)}\eta_{\mu\nu}e^{\mu}_{M}e^{\nu}_{N}
\approx e^{-2kb(x)}\eta_{\mu\nu}\delta^{\mu}_{M}\delta^{\nu}_{N}~, 
\end{eqnarray}
the kinetic energy term of gauge bosons remain as the canonical form 
and hence the gauge boson field $V_{\mu}$ is invariant 
under the above transformation,  
\begin{equation}\label{vectrf}
V_{\mu} \longrightarrow V_{\mu}~~.
\end{equation}
On the other hand, for the fermion field, we should take 
\begin{equation} \label{fermitrf}
\psi\longrightarrow \left(t\over \langle\phi\rangle\right)^{3/2}\psi 
\end{equation}
to make its kinetic term be canonical form.   
After redefinition of the fields, the original kinetic terms become  
\begin{equation}
{\cal L}~\supset -\frac{1}{4}F_{\mu\nu}F^{\mu\nu} 
+\left(1+\varphi/\langle\phi\rangle\right)^3
i~\overline{\psi}\gamma^{\mu}\partial_{\mu}\psi~~.
\end{equation} 

Next, let us consider the mass and general gauge interaction terms.   
With Eqs.~(\ref{4dscalar}), (\ref{vierbeintrf}), (\ref{vectrf}) 
and (\ref{fermitrf}), we obtain 
\begin{eqnarray}
-{\cal L}~~&&\supset e\bigg[-\frac{1}{2}M_V^2~V_MV^M+m_f\overline{\psi}\psi
+g\overline{\psi}\gamma^{\mu}e^{M}_{\mu}\psi V_{M}\bigg]
\nonumber \\
&& \longrightarrow  
~-\frac{1}{2}\bigg(1+\varphi/\langle\phi\rangle\bigg)^2
\bigg(M_V\langle\phi\rangle/t\bigg)^2~V_{\mu}V^{\mu}
+\bigg(1+\varphi/\langle\phi\rangle\bigg)^4
\bigg(m_f\langle\phi\rangle/t\bigg)~\overline{\psi}\psi \\
&&~~~~~~+\bigg(1+\varphi/\langle\phi\rangle\bigg)^3
~g\overline{\psi}\gamma^{\mu}\psi V_{\mu} \\
&&~~~= -\frac{1}{2}\bigg(1+2\varphi/\langle\phi\rangle\bigg)
\bigg(M_V\langle\phi\rangle/t\bigg)^2~V_{\mu}V^{\mu}
+\bigg(1+4\varphi/\langle\phi\rangle\bigg)
\bigg(m_f\langle\phi\rangle/t\bigg)~\overline{\psi}\psi \label{massint}\\
&&~~~~~~+\bigg(1+3\varphi/\langle\phi\rangle\bigg)
~g\overline{\psi}\gamma^{\mu}\psi V_{\mu} +\cdots ~~.\label{gaugeint}
\end{eqnarray}
If the parameters $M_V$ and $m_f$ are the 
Planck scales and $\langle\phi\rangle$ is TeV one, 
we could get TeV scale vector boson and fermion masses.  
We can show also that a scalar mass is suppressed by the factor 
$\langle\phi\rangle/t$ through the similar procedure \cite{rs}, 
which scenario is suggested by Randall and Sundrum first 
as a possible solution to the gauge hierarchy problem.  
However, from Eqs.~(\ref{massint}) and (\ref{gaugeint}), 
we can see that a low scale $\langle\phi\rangle$ possibly affects 
electroweak physics significantly by the radion interactions 
which in turn gives constraints on the mass of the radion.   
 
The low scale $\langle\phi\rangle$ can have observable 
effects at low energy. The dominant contribution comes at loop orders.
Since this theory is not renormalizable, we introduce
a cutoff $\Lambda$ in the Feynman loop integral.
If the soft supersymmetry breaking is introduced,
$\Lambda$ can be interpreted as the soft breaking scale.
However, the original Randall-Sundrum scenario for a
gauge hierarchy solution does not need supersymmetry and
hence our $\Lambda$ is not necessarily linked to
the soft supersymmetry breaking scale. Therefore, we simply
treat $\Lambda$ as an input parameter in this paper.

The loop integral generally involves the particle masses
in the loop. Hence, if W and Z bosons are in the loop,
their ratio is not proportional to their tree level value. 
The mass ratio of W and Z bosons determines the weak mixing angle.
Experimentally, the weak mixing angle is measured very precisely 
with errors of order $O(10^{-3})$. Also, the radion can contribute
to the $\Delta S=2$ process, but it turns out to be negligible. 
Figs.~1--4 show the typical contributions of the radion in some
relevant low energy processes.   

The Fermi constant is defined by the muon decay rate, 
$\mu^-\rightarrow e^-\nu_{\mu}\bar{\nu}_{e}$, and the weak 
Z boson coupling is determined by the weak neutral current
experiments such as $\nu_e e^-\rightarrow \nu_e e^-$.
With the radion, viz. Figs.~1 and 2, the couplings of the charged ($W$) 
and neutral ($Z$) gauge bosons are modified. Then, one can consider a
new mass parameter,
\begin{equation}\label{corrmass}
M_{W,Z}^{'2}=\frac{M_{W,Z}^2}{1+(\alpha_{W,Z}/\langle\phi\rangle^2 )}~
\end{equation}
where $\alpha_{W,Z}$ denotes the corrections by the radion intercations 
\begin{eqnarray}
\alpha_{W,Z}&\equiv&\frac{9M_{W,Z}^2}{i}\int \frac{d^4k}{(2\pi)^4}
\frac{1}{k^2-M_{W,Z}^2}\frac{1}{k^2-m_{\phi}^2}\nonumber \\
&=&\frac{9M_{W,Z}^2}{16\pi^2}
\int^1_0dx\bigg[\ln \left(\frac{\Lambda^2+M_{W,Z}^2+x(m_{\phi}^2-M_{W,Z}^2)}
{M_{W,Z}^2+x(m_{\phi}^2-M_{W,Z}^2)}\right)
-\frac{\Lambda^2}{\Lambda^2+M_{W,Z}^2+x(m_{\phi}^2-M_{W,Z}^2)}\bigg]\nonumber\\
&=&\frac{9M_{W,Z}^2}{16\pi^2}
\bigg[\frac{m_{\phi}^2}{m_{\phi}^2-M_{W,Z}^2}
\ln \left(1+\frac{\Lambda^2}{m_{\phi}^2}\right)
-\frac{M_{W,Z}^2}{m_{\phi}^2-M_{W,Z}^2}
\ln \left(1+\frac{\Lambda^2}{M_{W,Z}^2}\right)\bigg]~~.
\end{eqnarray}
where $\Lambda$ and $m_{\phi}$ denote the cuttoff scale and the radion mass, 
respectively. 
Of course, $\alpha_{W,Z}$ must be smaller than $\langle\phi\rangle^2$ 
in Eq.~(\ref{corrmass}) to maintain the validity of perturbativion.   
Note that {\it $\alpha_{W,Z}$ are monotonically decreasing functions 
as $m_{\phi}$ increase and $\alpha_Z>\alpha_W>0$ always}.  
Especially, {\it in the  $m_{\phi}\gg M_{W,Z}$ limit, they decrease very slowly 
because they become purely logarithmic functions in that limit}.
Note that a small $\alpha_{W,Z}$ is possible in the region
$\langle\phi\rangle^2\gg M_{W,Z}^2$, 
$m_{\phi}^2\gg \Lambda^2, M_{W,Z}^2$ or in the
region $\Lambda^2\ll m_{\phi}, M_{W,Z}^2$. 
But the last case makes our ``effective'' theory inconsistent 
because particles heavier than a cutoff scale are decoupled.

We can neglect the tree level contribution of
the radion (Fig.~3), since the couplings are 
given by $m_f/\langle\phi\rangle$ 
(viz. Eq.~(\ref{massint})). Especially, they are irrelevant 
in the experiments with {\it light} leptons.    

The modified effective masses of the gauge bosons could affect 
the determination of the weak mixing angle.
With the above relations, Eqs.~(\ref{corrmass}), 
the weak mixing angle $\theta_{w}^{'}$ is given by 
\begin{equation}
\cos^2 \theta_{w}^{'}\equiv \frac{M_{W}^{'2}}{M_{Z}^{'2}}
=\frac{M_{W}^{2}}{M_{Z}^{2}}\cdot 
\frac{1+\alpha_Z/\langle\phi\rangle^2}{1+\alpha_W/\langle\phi\rangle^2}
=\cos^2 \theta_{w}
\frac{1+\alpha_Z/\langle\phi\rangle^2}{1+\alpha_W/\langle\phi\rangle^2}~, 
\end{equation}
It can be expressed as
\begin{equation}
\arrowvert \frac{\Delta \sin^2 \theta_{w}}{1-\sin^2 \theta_{w}}\arrowvert
\ =\ \frac{\alpha_Z-\alpha_W}{\langle\phi\rangle^2+\alpha_W} < \delta\
,~~~{\rm or}~~~ 
\langle\phi\rangle^2 \ge\frac{1}{\delta}
\bigg(\alpha_Z-\alpha_W\bigg)-\alpha_W~
\end{equation}
where $\Delta \sin^2 \theta_{w}$ means 
($\sin^2 \theta_{w}^{'}-\sin^2 \theta_{w}$) and 
$\delta$ is the uncertainty.
Since $\alpha_{W,Z}$ are very slowly decreasing functions
in the region $m_\phi\gg M_{W,Z}$, the bound we obtain
is valid over a large region.

In Figs.~6 and 7, we show the allowed regions of
$\langle\phi\rangle$ and $m_\phi$ for $\Lambda=1,10$ TeV, respectively.
Here, we take the recent world average of
$\sin^2\theta_w=0.23124\pm 0.00024$~\cite{particle}.
From these, we note that for $\langle\phi\rangle\simeq 1-1.5$ TeV the weak 
mixing angle does not give a strong constraint at present.
However, for low values of $\langle\phi\rangle$ around
$<500$ GeV, the present value of the uncertainty in the
weak mixing angle gives a strong constraint. In these figures,
the shaded region is excluded by the 2 standard deviations
and the dotted line corresponds to the boundary of one
standard deviation.

There can be the radion contributions to the other parameters
of the low energy phenomena such as 
$(g-2)_{\mu}$ and $\Delta S=2$ processes. We comment
on these briefly. The present difference between theoretical
value and experimental value is negligible, 
$\frac{1}{2}(g-2)^{exp}_{\mu}-\frac{1}{2}(g-2)^{SM}_{\mu}=
(50.5\pm 46.5)\times 10^{-10}$~\cite{anom}.  
The radion contribution to $(g-2)_\mu$ is shown in Fig.~5, from
which we estimate
\begin{equation}
\frac{1}{2}(g-2)_\mu=\frac{m_{\mu}^4}{4\pi^2\langle\phi\rangle^2m_{\phi}^2}
\end{equation}
for $m_{\phi}\gg m_{\mu}$, and
\begin{equation}
\frac{1}{2}(g-2)_\mu=\frac{m_{\mu}^2}{4\pi^2\langle\phi\rangle^2}
\end{equation}
for $m_{\phi}\ll m_{\mu}$. 
From these, we obtain $m_\phi\langle\phi\rangle\ge 
20$~GeV$^2$ for $m_\phi\gg m_\mu$, and $\langle\phi\rangle\ge
200$~GeV for $m_\phi\ll m_\mu$.
We note that the radion contribution is negligible in $(g-2)_\mu$.

The radion can contribute to the $\Delta S=2$ process also at the
two loop level as shown in Fig.~5. Our estimate contributed by the
radion is
\begin{equation}
iT(\bar{K^0}\rightarrow K^0)=iT_{SM}
\left[1+\frac{1}{16\pi^2}\frac{M_W^2}{\langle\phi\rangle^2}
\times ({\rm log.\ div.\ terms})\right]
\end{equation}
where $T_{SM}$ is the SM contribution.
The data on $2{\rm Re}T$ is $(3.489\pm 0.009)\times 10^{-12}
$~MeV~\cite{particle}.
Thus, we obtain $\langle\phi\rangle\ge 160$~GeV.

In conclusion, it is pointed out that for a low cutoff scale $\Lambda$
the most significant radion effects can be
found in the weak mixing angle if the vacuum 
expectation value and mass of the
radion are below TeV scale. 

\acknowledgments
This work is supported in part by the BK21 program of Ministry 
of Education and Research grants through Center for
Theoretical Physics.

\newpage 
\begin{figure}
\epsfxsize=80mm
\centerline{\epsfbox{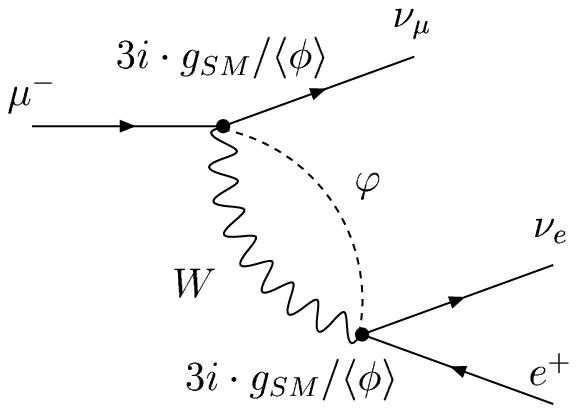}}
\end{figure}
\noindent Fig.~1. 
The muon decay diagram corrected by the radion.  
$g_{SM}$'s are the corresponding Standard 
Model coupling constants in the absence of the radion mediation. We omit
the subscripts in $g_{SM}$.   
\\\\\\

\begin{figure}
\epsfxsize=160mm
\centerline{\epsfbox{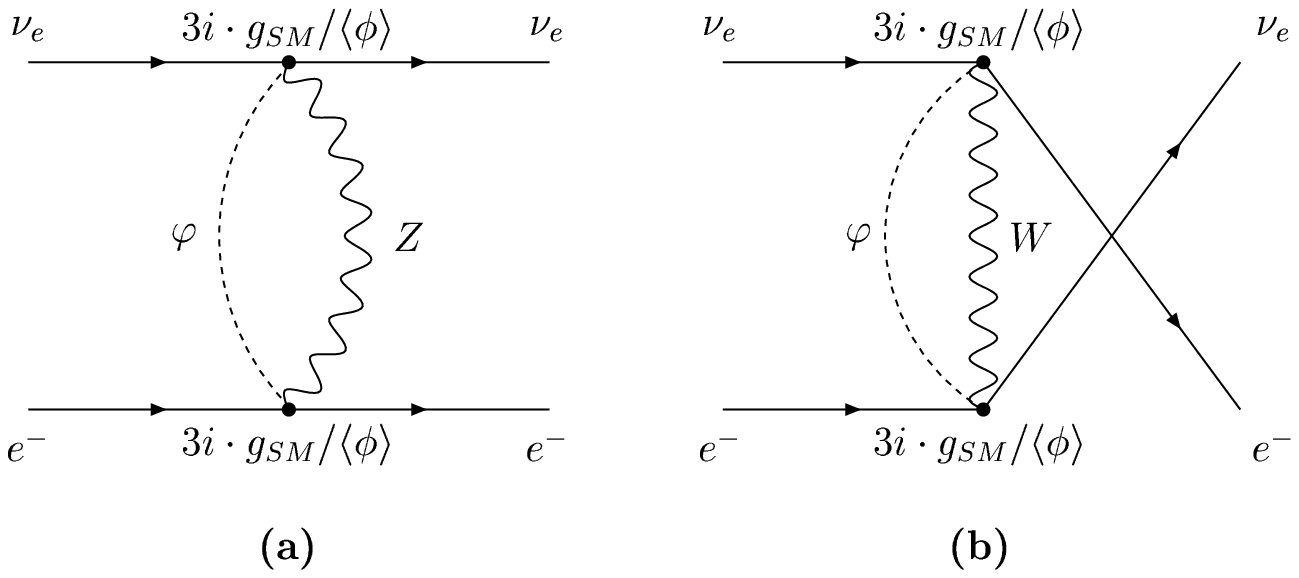}}
\end{figure}
\noindent Fig.~2.
The radion's contributions to 
$\nu_ee^-\rightarrow \nu_e e^-$.   
(a) and (b) corresponds to the neutral and charged 
currents, respectively.  
$g_{SM}$'s denote the corresponding Standard 
Model coupling constants in the absence of the radion mediation.

\newpage
\begin{figure}
\epsfxsize=80mm
\centerline{\epsfbox{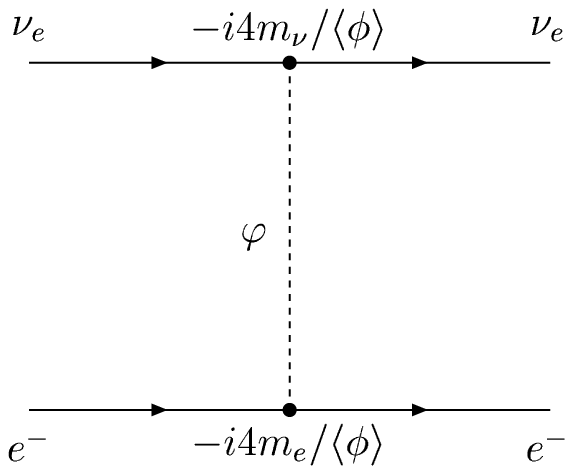}}
\end{figure}
\noindent Fig.~3. 
The radion tree level contribution to $\nu_e e^-\rightarrow\nu_e e^-$.
It is negligible because of the small couplings.
\\\\

\begin{figure}
\epsfxsize=160mm
\centerline{\epsfbox{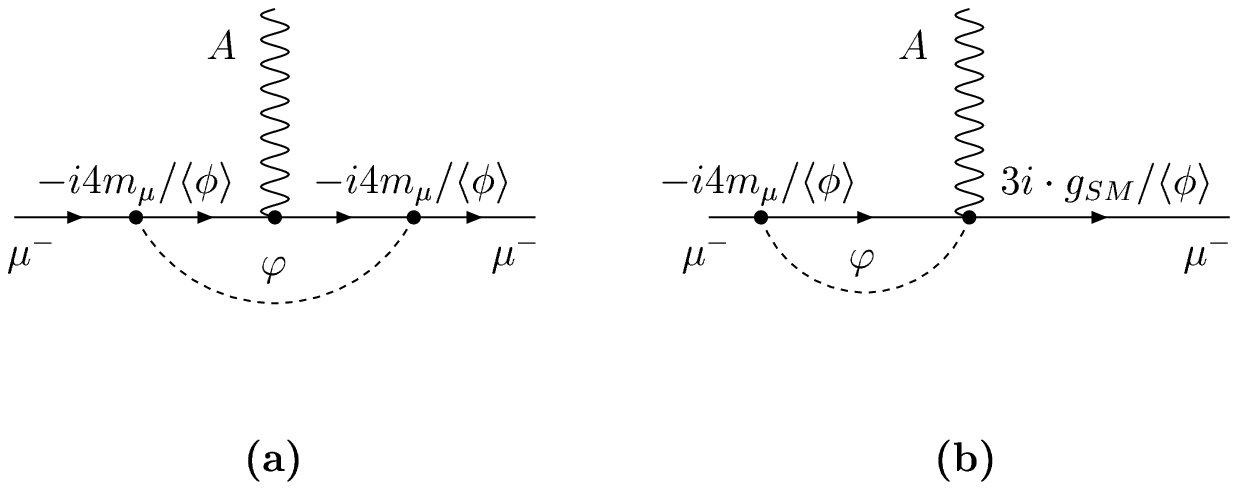}}
\end{figure}
\noindent Fig.~4.
The muon magnetic moment generated by the radion. 

\newpage
\begin{figure}
\epsfxsize=160mm
\centerline{\epsfbox{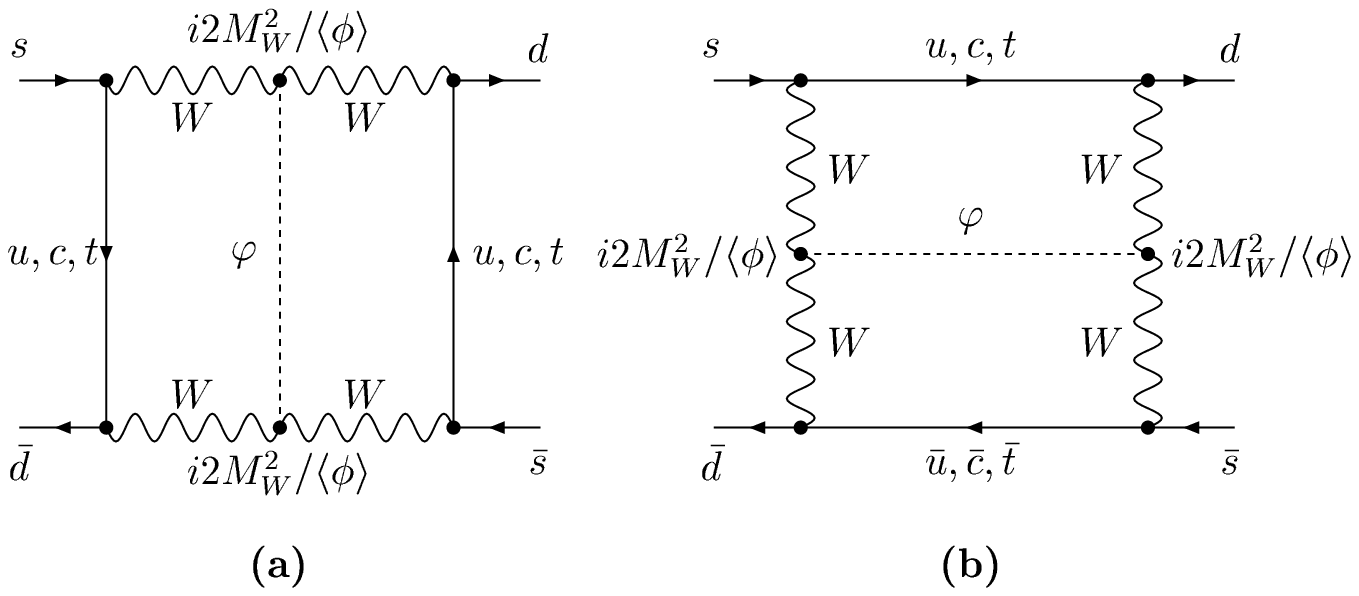}}
\end{figure}
\noindent Fig.~5. 
Dominant radion contributions to the $\Delta S=2$ process.

\newpage 
\begin{figure}
\epsfxsize=160mm
\centerline{\epsfbox{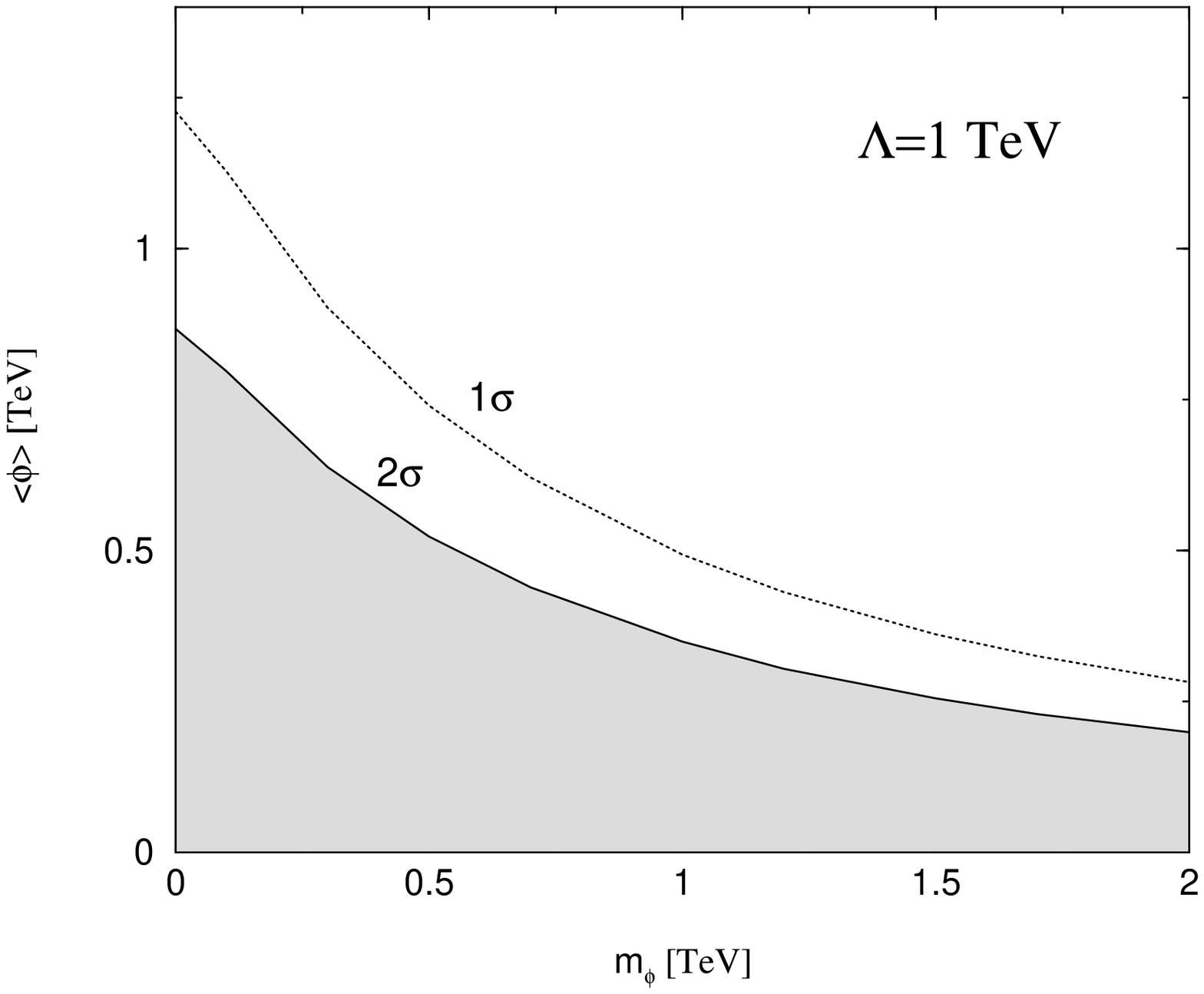}}
\end{figure}
\noindent Fig.~6.
The allowed region for the vacuum expectation value and mass
of the radion from the neutral current data. The cutoff
scale is 1~TeV.  

\newpage 
\begin{figure}
\epsfxsize=160mm
\centerline{\epsfbox{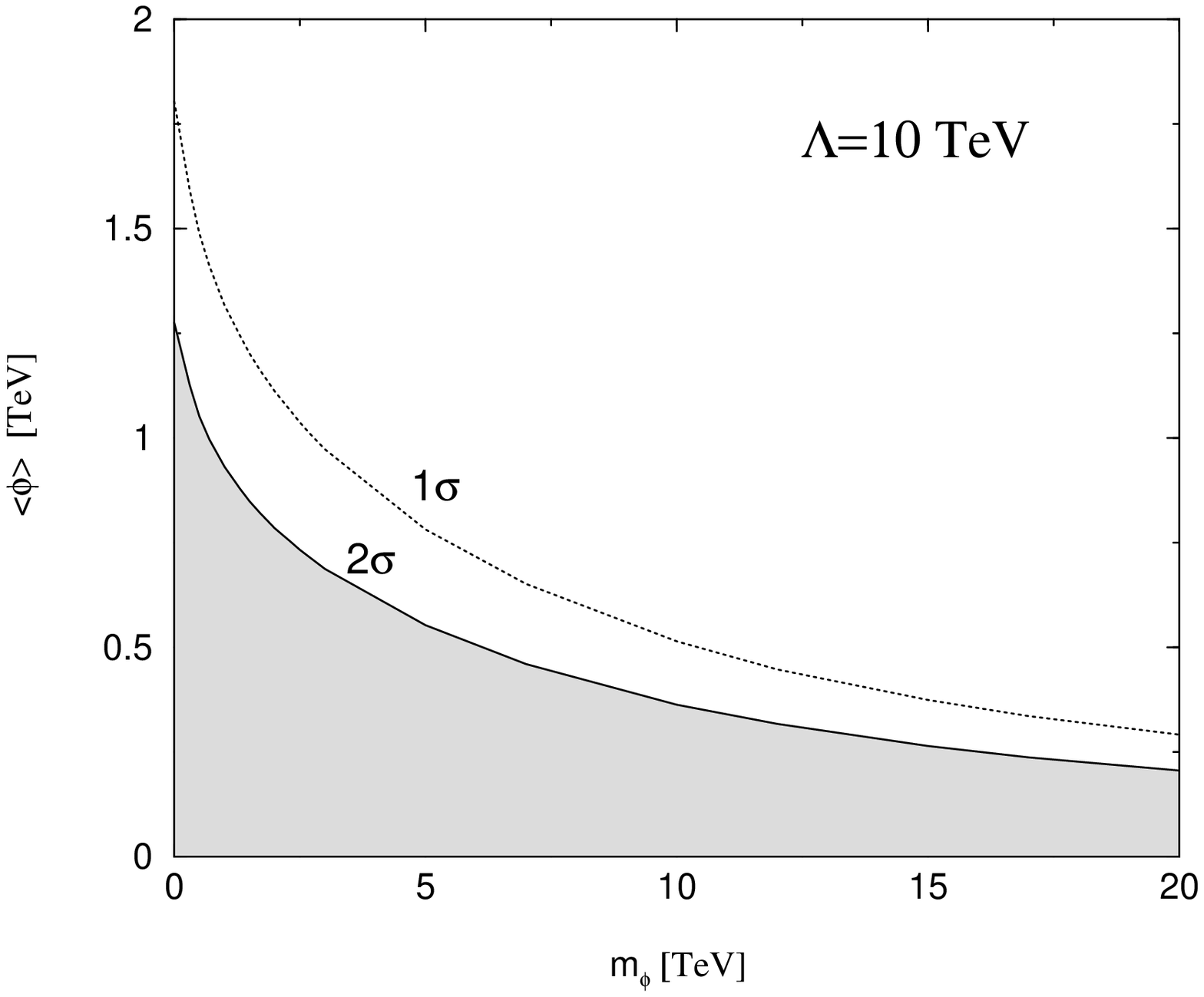}}
\end{figure}
\noindent Fig.~7.
The allowed region for the vacuum expectation value and mass
of the radion from the neutral current data. The cutoff
scale is 10~TeV.

\end{document}